\documentclass[12pt,amssymb,epsfig]{article}
\newcommand{\SC}{\scriptscriptstyle} 
 
\usepackage{graphicx}
\begin{document}

\title{\bf Observables in quantum gravity}
\author{{\bf Alejandro Perez\footnote{e-mail: perez@cpt.univ-mrs.fr}, 
Carlo Rovelli\footnote{e-mail: carlo@rovelli.org}}\\ 
\em CPT, Case 907, F-13288 Marseille, France;\\
\em Department of Physics, University of Pittsburgh, PA15260, USA.}
\date{\today}
\maketitle

\begin{abstract}
We study a family of physical observable quantities in quantum
gravity.  We denote them $W$ functions, or $n$-net functions.  They
represent transition amplitudes between quantum states of the
geometry, are analogous to the $n$-point functions in quantum field
theory, but depend on spin networks with $n$ connected components.  In
particular, they include the three-geometry to three-geometry
transition amplitude.  The $W$ functions are scalar under
four-dimensional diffeomorphism, and fully gauge invariant.  They
capture the physical content of the quantum gravitational theory.  

We show that $W$ functions are the natural $n$-point functions of the
field theoretical formulation of the gravitational spin foam models. 
They can be computed from a perturbation expansion, which can be
interpreted as a sum-over-four-geometries.  Therefore the $W$
functions bridge between the canonical (loop) and the covariant
(spinfoam) formulations of quantum gravity.  Following Wightman, the
physical Hilbert space of the theory can be reconstructed from the $W$
functions, if a suitable positivity condition is satisfied.

We compute explicitly the $W$ functions in a ``free'' model in which
the interaction giving the gravitational vertex is shut off, and we
show that, in this simple case, we have positivity, the physical
Hilbert space of the theory can be constructed explicitly and the
theory admits a well defined interpretation in terms of diffeomorphism
invariant transition amplitudes between quantized geometries.
\end{abstract}

\section{Introduction}

One of the hard problems in non-perturbative quantum gravity
\cite{india} is to construct a full set of physically meaningful
observable quantities \cite{observables}.  In this paper, we point out
that there is a natural set of quantities that one can define in
quantum general relativity, which are gauge invariant, have a natural
physical interpretation, and could play the role played by the
$n$-point functions in quantum field theory.  We denote these quantity
as $W$ functions, or $n$-net functions.  As the $n$-point functions in
a quantum field theory, these quantities are not natural quantities of
the corresponding classical field theory, namely in general
relativity.  Nevertheless, they capture the physical content of the
quantum theory and are related to the classical theory.

The $W$ functions are strictly related to the three-geometry to
three-geometry transition amplitude studied by Hawking \cite{haw}. 
However, they are not transition amplitudes between states in which
the classical three-geometry has an arbitrary sharp value, but rather
transition amplitudes between {\em eigenstates} of the three-geometry. 
In loop quantum gravity \cite{loop}, these eigenstates are
characterized by discretized geometries and are labelled by abstract
spin networks \cite{spinnet,carlolee}, or $s$-knots.  Thus the $W$
functions are rather transition amplitudes between states with fixed
amounts of ``quanta of geometry''.  This is analogous to the $n$-point
functions in field theory, which are not transition amplitudes between
field configurations, but rather transition amplitudes between states
characterized by a fixed number of ``quanta of field'' -- that is,
particles.  Furthermore, the $W$ functions generalize the
three-geometry to three-geometry amplitude (a 2-point function) to
arbitrary $n$-point functions; more precisely, we define the $W$
functions as a functional $W(s)$ over an {\em algebra\/} $\cal A$ of
abstract (not necessarily connected) spin networks.  In this respect,
the $W$ functions are analogous to the Wightman distributions \cite{W}
(hence the choice of the letter $W$).

We start from a general definition of the $W$ functions, based on
canonical quantum gravity.  We show that the $W$ functions are well
defined diffeomorphism invariant observable quantities and we clarify
their physical interpretation.  In this paper we focus on the case in
which the dynamics is ``real'', in a sense defined below.  The
physical meaning of this reality and the extension of the formalism to
the general case are discussed at the end of the paper.

A crucial property of the Wightman functions is the possibility of
reconstructing the quantum field theory from them -- a subtle
application of the beautiful Gelfand-Naimark-Segel (GNS)
representation theorem in the theory of $C^{*}$ algebras.  We show
here that the $W$ functions have the same property: under appropriate
conditions --in particular, a positivity condition-- the physical
Hilbert space of the theory and a suitable operator algebra can can be
reconstructed from the $W$ functions, using the GNS construction.  In
other words, we explore the extension to the generally covariant
context of Wightman's remarkable intuition that the content of a
quantum field theory is coded in its $n$-point functions.  To this
aim, we need to strip Wightman's theory from all the ``details'' that
follow from the existence of a background Minkowski space (positivity
of the energy, uniqueness of the vacuum, micro-locality\ldots) and show
that the core idea remains valid even in the absence of a background
spacetime.  For a line of investigation similar in spirit, see
\cite{abhay}.

Since the diffeomorphism invariant quantum field theory can be
characterized by its $W$ functions, the way is open for defining a
quantum theory of gravity by directly constructing its $W$ functions. 
Remarkably, spin foam models \cite{spinfoam,CarloMike} provide a
natural perturbative definition of $W$ functions.  In particular, it
has been recently shown that general spin foam models can be obtained
as the Feynman expansion of certain peculiar field theories over a
group \cite{dfkr,cm,ac}.  We show here that the gauge invariant
$n$-point functions of these field theories are precisely $W$
functions.  This construction provides a direct link between the field
theoretical formulation of the spin foam models and canonical quantum
gravity.  The link is similar in spirit to the link between the
operator definition quantum field theory and the construction of its
$n$-point functions via a functional integral \cite{W}.  In
particular, given the field theoretical formulation of a spin foam
model, we can construct a quantum gravity physical Hilbert space of
from its $W$ functions.  On the one hand, loop quantum gravity
provides the general framework and, in particular, the physical
interpretation of the $W$ functions; on the other hand, the field
theory over the group provides an indirect but complete definition of
the dynamics.  This is especially interesting on the light, in
particular, of the recent construction of {\em Lorentzian\/} spin foam
models \cite{BarrettCrane,Lorentzian}.  In turn, the perturbative
expansion of the field theory defines a sum over spinfoams which can
be directly interpreted as a sum over 4-geometries formulation of
quantum gravity.

A very interesting paper by Alexandar Mikovic \cite{mikovic} has
recently appeared, in which some of the idea presented here are
independently derived.  We present some comments on the relation
between the two approaches in the conclusion section.

We give an example of reconstruction in Section \ref{free}.  We
consider a simple ``free'' model, obtained by dropping the interaction
term which gives the quantum gravity vertex.  We prove positivity for
this case, and thus the existence of a Hilbert space of spin networks
for this quantum theory.  Finally, in Section \ref{openissues} we
discuss the meaning of the reality assumption and the extension to 
the complex case.

All together, we obtain an attracting unified picture, in which 
canonical loop quantum gravity, covariant spin foam models, and a 
family of diffeomorphism invariant physical observables for 
quantum gravity fit into a unified scheme. 

The ideas described in this paper have been first presented in the
second conference on Quantum Gravity in Warsaw, in June 1999.

\section{The 2-net function $W(s,s')$ in canonical quantum gravity}

In loop quantum gravity \cite{loop}, the Hilbert space ${\cal
H}_{diff}$ of the states invariant under three-dimensional
diffeomorphisms admits a discrete \cite{cedric} basis of states
$|s\rangle$, labelled by abstract spin networks (or $s$-knots) $s$.  An
abstract spin network $s$ is an abstract graph (not necessarily
connected) $\Gamma_{s}$, with links labelled by $SU(2)$
representations and nodes labelled by $SU(2)$ intertwiners (and
satisfying suitable conditions \cite{aj,cedric}).  There is a natural
union operation $\cup$ defined on the abstract spin networks: $s\cup
s'$ is the spin network defined by the the graph formed by the two
disconnected components $\Gamma_{s}$ and $\Gamma_{s'}$.

Notice that the relation between ${\cal H}_{diff}$ and the extended
Hilbert space ${\cal H}_{ex}$ formed by the unconstrained states
\cite{report,aj} can be expressed in terms of a ``projection" operator
$P_{diff}:{\cal H}_{ex}\mapsto{\cal H}_{diff}$ which sends an embedded
spin network state of a suitable basis of ${\cal H}_{ex}$ into a an
abstract spin network state, or $s$-knot, in ${\cal H}_{diff}$
\cite{CarloMike}.  Due to the infinite volume of the group of the
diffeomorphisms (or to the fact that the zero eigenvalue of the
diffeomorphism constraint is in the continuum spectrum) ${\cal
H}_{diff}$ is not a proper subspace of ${\cal H}_{ex}$ and $P_{diff}$
is not a true projection operator (hence the quotation marks), but
many alternative techniques for taking care of these technicalities
are known, and the space ${\cal H}_{diff}$ and the operator $P_{diff}$
are nevertheless well defined.

The states $|s\rangle$ have a straightforward physical interpretation,
which follows from the fact that they are projections on ${\cal
H}_{diff}$ of eigenstates of the area and volume operators
\cite{spinnet,carlolee}.  The interpretation is the following.  A
state $|s\rangle$ represents a three-geometry.  A three-geometry is an
equivalence class of three-metrics under diffeomorphisms.  The
geometry represented by $|s\rangle$ is quantized, in the sense that it
is formed by regions and surfaces having quantized values of volume
and area.  Intuitively, each node of $s$ represents a ``chunk'' of
space, whose (quantized) volume is determined by the intertwiner
associated to the node.  Two of such chunks of space are adjacent if
there is a link between the corresponding nodes.  The (quantized) area
of the surface that separates them is determined by the representation
$j$ associated to this link, according to the now well known relation
\cite{carlolee}
\begin{equation} 
    A = 8\pi \gamma \hbar G \ \sqrt{j(j+1)}
    \label{eq:area} 
\end{equation} 
where $\hbar, G, \gamma$ are the reduced Planck constant, the Newton
constant and the Immirzi parameter (the dimensionless free parameter
in the theory).  This interpretation of the states $|s\rangle$ follows
from the study of the area and volume operator on the Hilbert space of
the non-diffeomorphism invariant states.  Notice that the states
$|s\rangle$ are not gauge invariant either, and do not represent
physical gauge invariant notions.  The same is true for the
corresponding classical notion of three-geometry: a three-geometry is
determined by an ADM surface, which is a non-gauge-invariant notion in
general relativity.

The dynamics of the theory is given by the Hamiltonian constraint
$H(x)$, which we assume here to be a symmetric operator.  The space of
the solutions of this constraint is the physical Hilbert space of the
theory $H_{ph}$.  Instead of using the Hamiltonian constraint, we can
work with the linear operator $P: H_{diff} \to H_{ph}$ that projects
on the Kernel of $H(x)$.  (A suitable extension of $H_{diff}$ to its
generalized states --or any other of the many techniques developed for
this purpose-- should be used in order to take care of the technical
complications in defining the Hilbert eigenspace corresponding to an
eigenvalue in the continuum spectrum.)  For more details on this
operator, and, in particular, a more precise definition as a
three-diffeomorphism invariant object, see \cite{CarloMike}.  Instead
of worrying about the explicit construction of $P$, we assume here
that the operator $P: H_{diff} \to H_{ph}$ is given, and we consider
the quantity
\begin{equation}
    W(s,s') := \langle s | P | s' \rangle. 
    \label{eq:W2}
\end{equation} 
We claim that this is a well-defined fully gauge invariant quantity,
which represents a physical observable in quantum gravity and has a
precise and well-understood physical interpretation.

The gauge invariance of $W(s,s')$ is immediate.  All the objects on
the r.h.s.\,of (\ref{eq:W2}) are invariant under three-dimensional
diffeomorphisms, therefore we need to check only invariance under time
reparametrizations.  An infinitesimal coordinate-time shift is
generated by the Hamiltonian constraint.  If we gauge transform
(say) the bra state $\langle s|$ we obtain 
\begin{equation}
    \delta W(s,s') = \langle H s | P | s' \rangle =\langle s  | H P | 
    s' \rangle = 0,
    \label{eq:deltaW}
\end{equation} 
because $P$ is precisely the projector on the Kernel of $H$.  
Therefore $W(s,s')$ represents a gauge-invariant transition 
amplitude.  In fact, this is precisely the physical 
three-geometry to three-geometry physical transition amplitude.

To clarify why the three-geometry to three-geometry transition 
amplitude is a physical gauge-invariant quantity, consider 
a simple analogy with a well known system.  Consider a 
free relativistic particle in three spatial dimensions.  Its 
physical description is given by its position $\vec x(t)$ at each 
time $t$.  To have explicit Lorentz invariance in the formalism, 
the dynamics can be represented as a constrained 
reparametrization invariant dynamical system, by promoting the 
time variable $t$ to the role of dynamical variable $x^{0}=t$, 
and introducing an unphysical parameter ``time'' $\tau$.  The 
dynamics is then entirely determined by the constraints 
$p^{2}-m^{2}=0$ and $p^{0}>0$.  The corresponding constraints in 
the quantum theory are the Klein Gordon equation and the 
restriction to its positive frequency solutions.  The Hilbert 
space ${\cal H_{ex}}$ of the unconstrained states is formed by 
the square integrable functions on Minkowski space.  The physical 
Hilbert space ${\cal H}_{Ph}$ of the physical states is formed by 
the positive frequency solutions of the Klein Gordon equation.  
There is a well defined projection operator $P$, which restricts 
any state in ${\cal H}_{Ph}$ (more precisely, in the extension of 
${\cal H}_{Ph}$ which includes its generalized states) to its 
mass shell, positive frequency, component.  Now, consider the 
(generalized) state $|\vec x,x^{0}\rangle$ in ${\cal H}$.  This 
is the eigenstate of both the position $\vec x$ and the time 
$x^{0}$ operators, which are well defined self-adjoint operators 
on ${\cal H}$.  The interpretation of $|\vec x,x^{0}\rangle$ is 
clear: it is a particle at the Minkowski spacetime point $(\vec 
x,x^{0})$.  On the other hand, this is clearly {\em not\/} a physical 
state: there is no physical particle that can ``stay'' in 
a single point of spacetime (where is it after a second?).  It is 
a state that does not satisfies the dynamics.  Notice also that 
in ${\cal H}$ two such states at two different points of Minkowski 
space are orthogonal.  However, given the state $|\vec 
x,x^{0}\rangle$ in ${\cal H}$, we can project it down to 
$H_{Ph}$ and define the {\em physical\/} state 
\begin{equation}
     |\vec x,x^{0}\rangle_{Ph} =  P\, |\vec x,x^{0}\rangle.
    \label{eq:Px}
\end{equation}
In momentum space, this amounts to restrict it to its mass shell 
positive frequency components.  In coordinate space, this amount 
to spread out the delta function to a full solution of the Klein 
Gordon equation, which --as its happens-- at time $x^{0}$ is 
concentrated around $\vec x$, but at other times is spread around 
the future and past light cones of $(\vec x,x^{0})$.  The state 
$|\vec x,x^{0}\rangle_{Ph}$ is a physical state, and has a 
physical interpretation consistent with the dynamics: it is a 
(Heisenberg) state in which the particle is in $\vec x$ at time 
$x^{0}$, and has appropriately moved around in space at other 
times.  The transition amplitude between two such states 
is a physically meaningful quantity.  Indeed, it is nothing else 
that the familiar propagator in Minkowski space. But notice that 
\begin{equation} 
    W(\vec x,x^{0};\vec x{}',x^{0}{}')
    = {}_{Ph}\langle\vec x,x^{0}|\vec x{}',x^{0}{}'\rangle_{ph}
    = \langle\vec x,x^{0}|P|\vec x{}',x^{0}{}'\rangle,  
    \label{eq:W2particle}
\end{equation}
Namely the propagator is nothing but the matrix element of the 
projector operator $P$ between the {\em unphysical\/} states 
$|\vec x,x^{0}\rangle$! 

It is clear that the structure illustrated is the precisely the same
as in quantum gravity.  A classical three-geometry is determined by
three degrees of freedom per space point.  Two of these correspond to
physical degrees of freedom of the gravitational field, in analogy
with the dependent variable $\vec x$ above.  The third is the
independent temporal variable, in analogy with the $x^{0}$ variable in
the example above.\footnote{Of course, there is no a priori physical
distinctions between the two sets.  This is because the dynamics of
general relativity is relational: it provides relations between
equal-footing quantities, not a preferred temporal variable.  The
advantage of the formalism we are considering here is that it does not
require such a distinction to be made.  It does not require to single
out a preferred time variable.} Therefore $s$, precisely as $(\vec
x,x^{0})$ includes the dependent as well as the independent (time)
variables.  The states $|s\rangle$ are quantum states concentrated at
a single three-geometry.  Precisely as the states $|\vec
x,x^{0}\rangle$, these are unphysical, because spacetime cannot be
concentrated on a unique three-geometry, in the very same sense in
which a particle cannot be at a unique point of Minkowski space.  The
projector $P$ project a state $|s\rangle$ into a physical state which
spreads across three-geometries, and the transition amplitude
(\ref{eq:W2}) gives the amplitude of measuring the three-geometry
corresponding to $s$ after we have measured the three geometry
corresponding to $s'$.  This amplitude is well defined and
diffeomorphism invariant.

\section{Reality of $P$ and $W$ functions} 
\label{Wfunctions}

Let us now return to the gravitational theory.  We assume in this 
section that $P$ has the following property, which we call (for 
reason that will become clear later on) ``reality''
\begin{equation}
       \langle s_{1}\cup s_{3} |P| s_{2} \rangle = 
       \langle s_{1} |P| s_{2} \cup s_{3} \rangle. 
    \label{eq:reality}
\end{equation}
The physical meaning of this property, as well as the extension 
of the formalism to the case in which this property does not hold 
are discussed in Section \ref{openissues}. 

Consider the vector in ${\cal H}_{ph}$
\begin{equation}
    |0 \rangle_{ph}\equiv P |0\rangle.
    \label{eq:0Ph}
\end{equation}
and, in general,
\begin{equation}
        |s \rangle_{ph}\equiv P |s\rangle.
    \label{eq:sPh}
\end{equation}
(See the particle analogy discussed at the end of last 
section.)  The 2-net function $W(s,s')$, defined in 
(\ref{eq:W2}), can then be written also as
\begin{equation}
    W(s,s') = {}_{ph}\langle s |s' \rangle_{ph}. 
\end{equation}
Clearly the states $|s \rangle_{ph}$ form an overcomplete basis of
${\cal H}_{ph}$.  In particular, there will be relations
between them, of the form
\begin{equation}
       \sum_{s} c_{s}|s \rangle_{ph}=0
    \label{eq:linearrel}
\end{equation}
for appropriate complex numbers $c_{s}$.  Notice that 
(\ref{eq:linearrel}) is equivalent to $P \sum_{s} 
c_{s}|s \rangle=0$, or $\langle s' P \sum_{s} c_{s}|s 
\rangle=0, \forall s'$.  This can also be rewritten as 
$\langle s'\cup s'' P \sum_{s} c_{s}|s \rangle=0, 
\forall s'$ and, because of the reality 
(\ref{eq:reality}) of $P$, as $\langle s' P \sum_{s} 
c_{s}|s\cup s" \rangle=0, \forall s'$.  Therefore
\begin{equation}
       \sum_{s} c_{s}|s\cup s'' \rangle_{ph}=0
    \label{eq:linearrel2}
\end{equation}
for all $s''$, whenever (\ref{eq:linearrel}) holds.  Using this fact,
we define on ${\cal H}_{ph}$ the operator
\begin{equation}
\hat\phi_{s}|s' \rangle_{ph}=|s'\cup s \rangle_{ph}.  
\label{phifull}
\end{equation}
This definition is well posed, in spite of the 
overcompleteness of the vectors $|s\rangle_{ph}$, 
because of (\ref{eq:linearrel2}), that is, $\phi_{s}$ 
sends the vanishing linear combinations 
(\ref{eq:linearrel}) of states into the linear 
combinations (\ref{eq:linearrel2}) which are still 
vanishing.  Also, notice that $\hat\phi_{s}$ is 
self-adjoint, again because of the reality of $P$
\begin{eqnarray}
 {}_{ph} \langle s_{1} | \phi_{s}^{\dagger}| s_{2} \rangle_{ph} &=&
 {}_{ph} \langle \phi_{s} s_{1}| s_{2} \rangle_{ph} = {}_{ph} \langle
 s_{1}\cup s| s_{2} \rangle_{ph} = \langle s_{1}\cup s |P| s_{2} \rangle
 \nonumber \\
&=& \langle s_{1} |P| s_{2} \cup s \rangle\ {}_{ph} \langle s_{1} |
s_{2}\cup s \rangle_{ph} = {}_{ph} \langle s_{1} | \phi_{s}| s_{2}
\rangle _{ph} ,	
\label{dagger=}
\end{eqnarray}
and  it commutes with itself
\begin{equation}
    [\hat\phi_{s},\hat\phi_{s'}]=0,
    \label{eq:commutativity}
\end{equation}
since 
\begin{equation}
    \hat\phi_{s}\hat\phi_{s'}=\hat\phi_{s'}\hat\phi_{s}=\hat\phi_{s\cup s'}.
    \label{eq:comm2}
\end{equation}

The 2-net function $W(s,s')$, defined in (\ref{eq:W2}), can be written
now as
\begin{equation}
    W(s,s') = {}_{ph}\langle 0|\hat\phi_{s}\hat\phi_{s'}|0 \rangle_{ph}. 
\label{nostar}
\end{equation}
More in general, we can define 
\begin{equation}
    W(s) = {}_{ph}\langle 0|\hat\phi_{s}|0 \rangle_{ph}. 
    \label{eq:W}
\end{equation}
so that 
\begin{equation}
    W(s,s')=W(s\cup s'). 
\end{equation}

Now, consider the free linear space $\cal A$ formed by the (formal)
linear combinations of spin networks, with complex coefficients 
\begin{equation}
    A = \sum_{s} c_{s} s. 
    \label{eq:v}
\end{equation}
There is a natural product defined on $\cal A$ by $s\cdot s'=s\cup
s'$, and a natural star operation defined by $s^{*}=s$ (Here we refer
to spin networks labeled by $SU(2)$ representations and each
representation of $SU(2)$ is conjugate to itself.  When spin networks
are labeled by representations of groups which are not self-conjugate
the star operation should replaces representations with dual
representations.)  We define the norm $||A||=sup_{s}|c_{s}|$.  We
obtain in this way a $C^{*}$ algebra structure on $\cal A$.  The
quantity $W(s)$, defined in (\ref{eq:W}), defines a linear functional
on $\cal A$.  A straightforward calculation shows that the functional
is positive
\begin{equation}
    W(A^{*}A)\ge 0.
    \label{eq:positivity}
\end{equation}
We can thus apply the Gelfand-Naimark-Segal construction to the
$C^*$algebra $\cal A$ and the positive linear functional $W$, obtaining
a Hilbert space $\cal H$, a ``vacuum'' state $|0 \rangle$ and a 
representation $\phi$ of $\cal A$ in the Hilbert space, such that 
\begin{equation}
    W(s) = \langle 0|\hat\phi(s)|0 \rangle. 
\end{equation}
But it is clear that in doing so we have simply reconstructed the
Hilbert space ${\cal H}_{ph}$, the ``vacuum'' state $|0 \rangle_{ph}$
and the algebra of the operators $\hat\phi_{s}$.  In other words, the
content of the canonical theory of quantum gravity can be coded, in
the spirit of Wightman, in the positive linear functional $W(s)$ over
the algebra $\cal A$ of the spin networks.

We can thus determine the dynamics of the theory by giving $W(s)$,
instead of explicitly giving the projector $P$, or the Hamiltonian
constraint, and reconstruct the physical Hilbert space from $W(s)$.  
In particular, the main physical gauge-invariant observable, namely 
the three-geometry to three-geometry transition amplitude is simply 
the value of $W(s)$ on the spin networks $s$ formed by two disjoint 
components. 

We close this section with a comment about locality. 
The sense in which general relativity is a local theory is far more
subtle that in ordinary field theory.  For a detailed discussion of
this issue see for instance \cite{observables}.  In particular, physical
gauge invariant observables are independent from the spacetime
coordinates $\vec x,t$, and therefore they are not localized on the
spacetime manifold, which is coordinatized by $\vec x,t$. 
Nevertheless, the dynamics of general relativity is still local in an
appropriate sense.  This locality should be reflected in a general
property of the $W$ functions.  Roughly, we expect that if a spin
network $s$ can be cut in two parts (connected to each each other)
$s_{ext}$ and $s_{in}$, and a second spin network $s'$ can be cut in
two parts (connected to each each other) $s'_{ext}$ and $s'_{in}$, and
if $s_{ext}=s'_{ext}$, then $W(s,s')$ should be independent from
$s_{ext}$.  In other words, the local evolution in apart of the spin
network should be independent from what happens elsewhere on the spin
network.  A precisely formulation of this property and its
consequences deserve to be studied.

\section{$W(s)$ in field theories over a group}

In the last few years, intriguing developments in quantum gravity have
been obtained using the spin foam \cite{spinfoam} formalism. 
Recently, it has been shown that any spin foam model can be derived
from an auxiliary field theory over a group manifold \cite{dfkr,cm}. 
Several spin foam models defined from auxiliary theories defined over
a group have been developed.  They are covariant, have remarkable
finiteness properties \cite{ac}, exist in Lorentzian form
\cite{Lorentzian} and represent intriguing covariant models for a
quantum theory of the gravitational field.  In this section, we
illustrate the emergence of a $W(s)$ functional over $\cal A$ in the
context of these field theories over a group manifolds.  For a similar
derivation see \cite{mikovic}.

For concreteness, and simplicity of the presentation, let us consider
a specific model.  Consider a real field theory for a scalar field
defined over a group manifold $\phi(g^{i}) =
\phi(g^{1},g^{2},g^{3},g^{4})$, where $g^{i} \in G$, which we chose
for the moment to be a compact Lie group \cite{dfkr,ac}.  The field
$\phi(g^1,g^2,g^3,g^4)$ is defined to be symmetric under permutation
of its four arguments and $G$ invariant in the sense that it satisfies
$P_g\phi=\phi$, where the operator $P_g$ is defined by
\begin{equation} 
\label{pg} P_{g}\phi(g^1,g^2,g^3,g^4) \equiv \int_{G} dg \
{\phi}(g^1g,g^2g,g^3g,g^4g),
\end{equation}
The dynamics is given by an action $S[\phi]$, which we do not specify 
for the moment. The $n$-point functions of the theory have the form 
\begin{equation}
W(g^{i_{1}}_{1},\ldots, g^{i_{n}}_{n}) = \int [D\phi]\
\phi(g^{i_{1}}_{1}), \ldots, \phi(g^{i_{n}}_{n})\ e^{iS[\phi]}.
\end{equation} 
Let us work in momentum space.  Using Peter-Weyl
theorem, we expand the field in terms of the matrix elements of the
irreducible representations $D^{(N)}_{\alpha\beta}$ of $G$. 
\begin{equation}
    \phi(g_1,\dots, g_4) =\sum \limits_{N_1\dots N_4} 
    \Phi^{\alpha_1 \dots \alpha_4}_{{\SC 
(N_1 \dots N_4)}\beta_1\dots\beta_4} \  
D^{(N_1)\beta_1}_{\alpha_1}(g_1) \dots 
D^{(N_4)\beta_4}_{\alpha_4}(g_4). 
    \label{eq:PW}
\end{equation}
We denote as $C^{N_1\ldots N_4\,\Lambda}_{\alpha_1\ldots \alpha_4}$ a
normalized basis in the space of the intertwiners between the
representations $N_1\ldots N_4$.  Imposing the $G$ invariance of the
field on the momentum space components, and using the relation
\begin{equation} \label{app:4-int}
 \int\limits_{G} dg \ D^{(N_1)}_{\alpha_1\beta_1}(g) 
 \ldots D^{(N_4)}_{\alpha_4\beta_4}(g) = \sum_\Lambda \ \ 
 C^{N_1\ldots N_4\,\Lambda}_{\alpha_1\ldots \alpha_4}\ \ 
 {C}^{N_1\ldots N_4\,\Lambda}_{\beta_1\ldots \beta_4}
\end{equation}
we can write the field as 
\begin{eqnarray}\label{cul}
&& \phi(g_1,\dots, g_4)=   \\ 
&& \ \ \ \ \sum \limits_{N_1\dots N_4} \Phi^{\alpha_1 
\dots \alpha_4}_{{\SC 
(N_1 \dots
N_4)}\beta_1\dots\beta_4} \  
D^{(N_1)\gamma_1}_{\alpha_1}(g_1) \dots 
D^{(N_4)\gamma_4}_{\alpha_4}(g_4)
\sum \limits_\Lambda 
C^{\SC N_1 \dots N_4,\Lambda}_{\gamma_1\dots 
\gamma_4} C_{\SC N_1 \dots N_4,\Lambda}^{\beta_1\dots\beta_4},
\nonumber
\end{eqnarray}
or, defining (for later convenience)  
\begin{equation}
\phi^{\alpha_1\ldots\alpha_4}_{\SC
N_1\ldots N_4,\Lambda} := {\Phi^{\alpha_1
\dots \alpha_4}_{{\SC  (N_1 \dots
N_4)}\beta_1\dots\beta_4} {C}^{N_1\ldots
N_4\,\Lambda}_{\beta_1\ldots \beta_4}\over
\Delta_{N_1}\Delta_{N_2}\Delta_{N_3}\Delta_{N_4}}, 
\end{equation} 
as 
\begin{eqnarray} 
&& \phi(g_1,\dots, g_4)= \\ \nonumber 
&& \ \ \ \sum\limits_{N_1\dots N_4, \Lambda} 
\! \phi^{\alpha_1\ldots\alpha_4}_{\SC
N_1\ldots N_4,\Lambda}
\left( \Delta_{N_1}\ldots\Delta_{N_4} 
D^{(N_1)\gamma_1}_{\alpha_1}(g_1) \dots 
D^{(N_4)\gamma_4}_{\alpha_4}(g_4)
C^{\SC N_1 \dots N_4,\Lambda}_{\gamma_1\dots 
\gamma_4}\right). 
\end{eqnarray}
We can take the quantities
$\phi^{\alpha_1\ldots\alpha_4}_{\SC
N_1\ldots N_4,\Lambda}$
as the independent ``Fourier components'' of the field, and therefore
write the $W$ functions, in momentum space as
\begin{equation}
    W^{\alpha^{1}_1\alpha^{1}_2\alpha^{1}_3\alpha^{1}_4}_{\SC
    N^{1}_1N^{1}_2N^{1}_3N^{1}_4,\Lambda^{1}}\ldots
{}^{\alpha^{n}_1\alpha^{n}_2\alpha^{n}_3\alpha^{n}_4}_{\SC
N^{n}_1N^{n}_2N^{n}_3N^{n}_4,\Lambda^{n}}       
= \int [D\phi]\ \phi^{\alpha^{1}_1\alpha^{1}_2\alpha^{1}_3\alpha^{1}_4}_{\SC
N^{1}_1N^{1}_2N^{1}_3N^{1}_4,\Lambda^{1}} 
\ldots \phi^{\alpha^{n}_1\alpha^{n}_2\alpha^{n}_3\alpha^{n}_4}_{\SC
N^{n}_1N^{n}_2N^{n}_3N^{n}_4,\Lambda^{n}}  \ \ 
e^{iS[\phi]}. 
\end{equation} 
However, the measure and the action are $G$ invariant.  Therefore the
only nontrivial independent $W$ functions are given by $G$ invariant
combinations of fields, where $G$ acts on each index $\alpha_{i}^n$ by
the representation $N_{i}^n$ .  There is only one way of obtaining $G$
singlets: to have the indices $\alpha_{i}^n$ all paired --with the
two indices of the pair sitting in the same representation-- and
to sum over the paired indices.  Each independent $W$ function is
determined by a choice of indices and their pairing.  

In order to describe these index choices and pairings, 
let us associate to
each field $\phi^{\alpha_1\alpha_2\alpha_3\alpha_4}_{\SC
N_1N_2N_3N_4,\Lambda}$ in the
integrand a four valent node; we associate to this node the
intertwiner $\Lambda^{n}$, and to each of its four links a 
representation $N^{n}_{i}$. We then connect the links between two 
fields with paired indices.  We obtain a graph, with nodes labelled by 
intertwiners and links labelled by representations (satisfying 
Clebsch-Gordan like relations), namely a spin network $s$ (in the 
group $G$). Thus, independent $W$ functions are labelled by spin 
networks!   

In other words, to each spin network $s$, with nodes $n$ 
labelled by intertwiners $\Lambda_{n}$ and links $l$ labelled by 
representations $N_{l}$, we can associate 
a gauge invariant product of field operators $\phi_{s}$
\begin{equation}
    \phi_{s} = \sum_{\alpha_{l}} \ 
    \prod_{n} \ \phi^{\alpha^{n}_1\alpha^{n}_2\alpha^{n}_3\alpha^{n}_4}_{\SC
N^{n}_1N^{n}_2N^{n}_3N^{n}_4,\Lambda^{n}}
\end{equation}
where $\alpha^{n}_i$ is the index associated to the 
link $l$ which is the $i$-th link of the node $n$.
And we define
\begin{equation}
    W(s) = \int [D\phi]\ \ \phi_{s} \ \ e^{iS[\phi]}. 
\end{equation} 

Therefore the field theory 
over the group defines a $W$ functional over the spin networks 
algebra $\cal A$. 
If $W(s)$ is positive, we have then immediately, thanks to the GNS
theorem, a Hilbert space and an algebra of field operators whose
vacuum expectation value is $W(s)$.  Under suitable conditions, this
could be identified with the physical Hilbert space of quantum
gravity.  For this, the group $G$ has to be $SU(2)$, or,
alternatively, the representations and the intertwiners should be in
correspondence with the ones of $SU(2)$.  This is the case in
particular for the gravitational $SO(4)$ and $SO(3,1)$ models 
\cite{BarrettCrane,Lorentzian} 
in which the dynamics restricts the representations to the
simple, or balanced, representations, which can be identified with
the irreducible $SU(2)$ representations.  

The relation between the field theory and quantum 
gravity becomes much more transparent by expressing 
$W(s)$ explicitly as a perturbation expansion.  Indeed, 
as shown in Ref.\,\cite{dfkr,cm,ac}, the standard field 
theoretical perturbation expansion of $W(s)$ in Feynman 
graphs turns out to be a sum over spin foams.  
In particular $W(s,s') = W(s\cup s')$ is given by a sum 
over all spin foams $\sigma$ bounded by the spin 
networks $s$ and $s'$
\begin{equation}
    W(s,s') = \sum_{\sigma,\ \partial\sigma=s\cup s'}\ A(\sigma). 
    \label{eq:sigma}
\end{equation}
A spin foam $\sigma$ admits an interpretation as a (discretized)
4-geometry.  In particular, $\sigma$ can be the complex dual to a four
dimensional cellular complex, and the representations and intertwiners
are naturally related to areas and volumes of the elementary 2 and 3
cells.  Therefore (\ref{eq:sigma}) is a (precise) implementation of
the representation of quantum gravity as a sum over geometries,
introduced by Wheeler and Misner \cite{misner}, and developed by
Hawking and collaborators \cite{haw}.  The generation of the
spacetimes summed over as Feynman diagrams is a four-dimensional
analog of the two-dimensional quantum gravity models developed
sometime ago in the context of the string theory in zero dimensions 
\cite{2d}.

\section{A ``free'' theory}\label{free}

As a simple example, we sketch here the structure of a very simple
model in which the action $S[\phi]$ contains only a kinetic part and
no interaction part.
\begin{equation}
    S[\phi]= i/2\ \int dg^1\ldots dg^4\ \phi^{2}(dg^1\ldots dg^4). 
    \label{eq:action}
\end{equation}
A straightforward calculation yields the action in momentum space. 
\begin{equation}
S[\phi] \!  \!=\!\!  \frac{i}{2}
\phi^{\alpha_1\ldots\alpha_4}_{\SC N_1\ldots N_4,\Lambda}
\phi^{\beta_1\ldots\beta_4}_{\SC M_1\ldots M_4,\tilde \Lambda}
\left({(\Delta_{N_1}\dots \Delta_{N_4})\delta_{\alpha_1\beta_1}
\ldots\delta_{\alpha_4\beta_4}
\delta^{\SC N_1M_1}\ldots\delta^{\SC
N_4M_4}\delta^{\SC \Lambda \tilde \Lambda} }\right),
\end{equation} 
where sum over repeated indices is understood.

Every $n$-point function of the field theory can be calculated as
functional derivatives of the generating function ${\cal W}(J)$
defined as:
\begin{equation}
{\cal W}(J)=\int {\cal D} \phi \ \ exp \left( i S[\phi] +
J^{\alpha_1\ldots\alpha_4}_{\SC N_1\ldots N_4, \Lambda} 
\phi_{\alpha_1\ldots \alpha_4}^{\SC N_1\ldots N_4, 
\Lambda}\right),
\end{equation}
which can be easily computed using standard Gaussian 
integration:  
\begin{equation}
W(J)={\cal C}\ exp \left({1 \over 2} 
{J^{\alpha_1\ldots\alpha_4}_{\SC
N_1\ldots N_4, \Lambda}\  J_{\alpha_1\ldots\alpha_4}^{\SC
N_1\ldots N_4, \Lambda}\over \Delta_{N_1}\dots
\Delta_{N_4}}\right), \end{equation}
The $n$-point function depends on the boundary 4-valent spin network
defined by the action of the $J^{\alpha_1\dots \alpha_4}_{\SC
N_1\ldots N_4, \Lambda}$'s which can be represented by a 4-valent node
carrying the representations $N_1$, $N_2$, $N_3$, $N_4$ respectively
and an intertwiner colored by $\Lambda$.  Contraction of their indices
$\alpha_i$ is represented by the connection of the corresponding links.
Lets illustrate with an example the  computation of the Wightman functions
for this free theory. The two point function $W(s_1,s_2)$ is given by 
\begin{equation}
W(s_1,s_2)=\left\{{\delta \over \delta
J^{\alpha_1\ldots\alpha_4}_{\SC N_1\ldots N_4, 
\Lambda}}{\delta \over
J^{\alpha_1\ldots\alpha_4}_{\SC N_1\ldots N_4, \Lambda}}
{\delta \over \delta J^{\beta_1\ldots\beta_4}_{\SC
M_1\ldots M_4, \Gamma}}{\delta \over \delta 
J^{\beta_1\ldots \beta_4}_{\SC
M_1\ldots M_4, \Gamma}} W(J) \right\}_{J=0}, 
\end{equation}
where the boundary spin networks $s_1$, and $s_2$ are given by the 
corresponding contraction of the $J$'s functional derivatives, namely:
\begin{eqnarray} \label{s1}
{\delta \over \delta
J^{\alpha_1\ldots\alpha_4}_{\SC N_1\ldots N_4, 
\Lambda}}{\delta \over
J^{\alpha_1\ldots\alpha_4}_{\SC N_1\ldots N_4, \Lambda}}
 \rightarrow \begin{array}{c} 
\includegraphics[width=4cm]{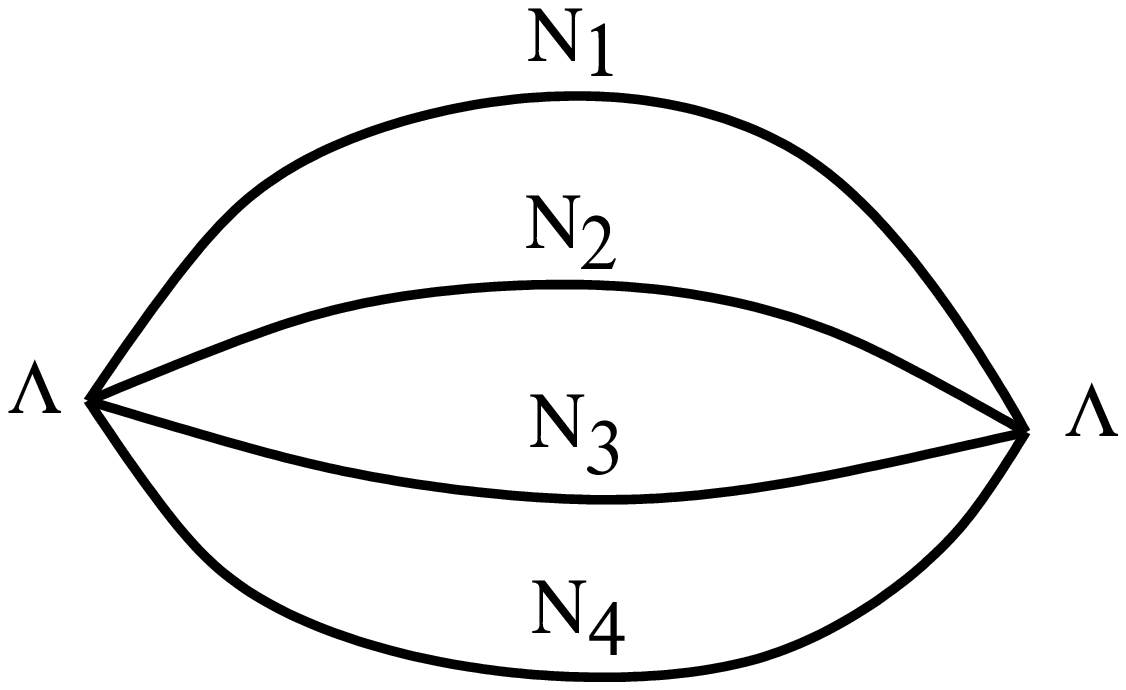}\end{array},
\end{eqnarray}
and
\begin{eqnarray}\label{s2}
{\delta \over \delta J^{\beta_1\ldots\beta_4}_{\SC
M_1\ldots M_4, \Gamma}}{\delta \over \delta 
J^{\beta_1\ldots \beta_4}_{\SC
M_1\ldots M_4, \Gamma}}
 \rightarrow \begin{array}{c} 
\includegraphics[width=4cm]{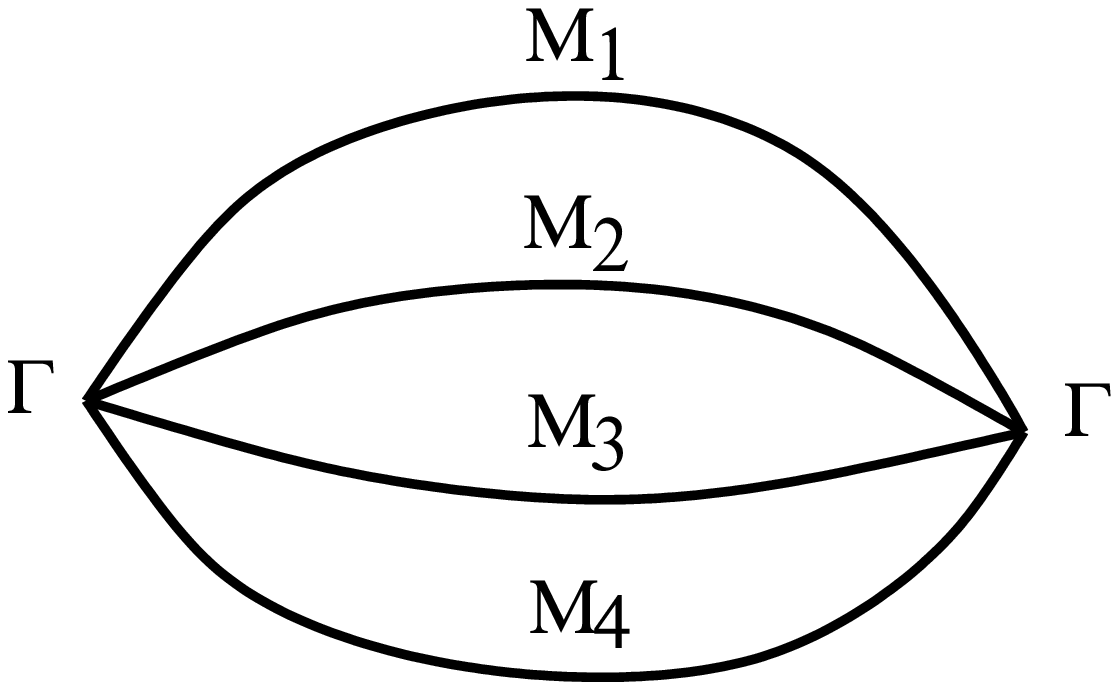}\end{array}. 
\end{eqnarray}
A straight forward calculation gives
\begin{equation}
W(S_1,S_2)= 1 +
\delta_{N_1M_1}\ldots \delta_{N_4M_4}\delta_{\Lambda
\tilde \Lambda}. \end{equation}

The $C^*$ algebra $\cal A$ is defined as the algebra as the free sum of
(non necessarily connected) 4-valent spin networks over $SO(4)$ as in (\ref{eq:v}). 
The $*$ operation is simply defined by complex conjugation of the 
components of $A$. We define the functional $W$ over the algebra by means of
the corresponding Wightman functions of our spin foam model. 
The fact that the positivity condition holds for $W$ (namely, $W({
A}^*{A}) \ge 0$) can be easily seen from the form of the functional measure.
We can explicitly construct an orthonormal basis in ${\cal H}_{ph}$ as follows.  
There are two kind of situations: spin network which do not interfere
with the vacuum $\left|0 \right>$ (the empty spin network), and those which do. 
In the first case the projection is trivial and the elements of 
the physical Hilbert space ${\cal H}_{ph}$ are the simply the 
original spin network states.
Some examples are the following 
\begin{eqnarray}
\nonumber
&&\left|0 \right>  = 1, 
\end{eqnarray}
\begin{eqnarray} 
\nonumber &&\left| ij, \lambda  \right>_{\begin{array}{c}
	\includegraphics[width=.25cm]{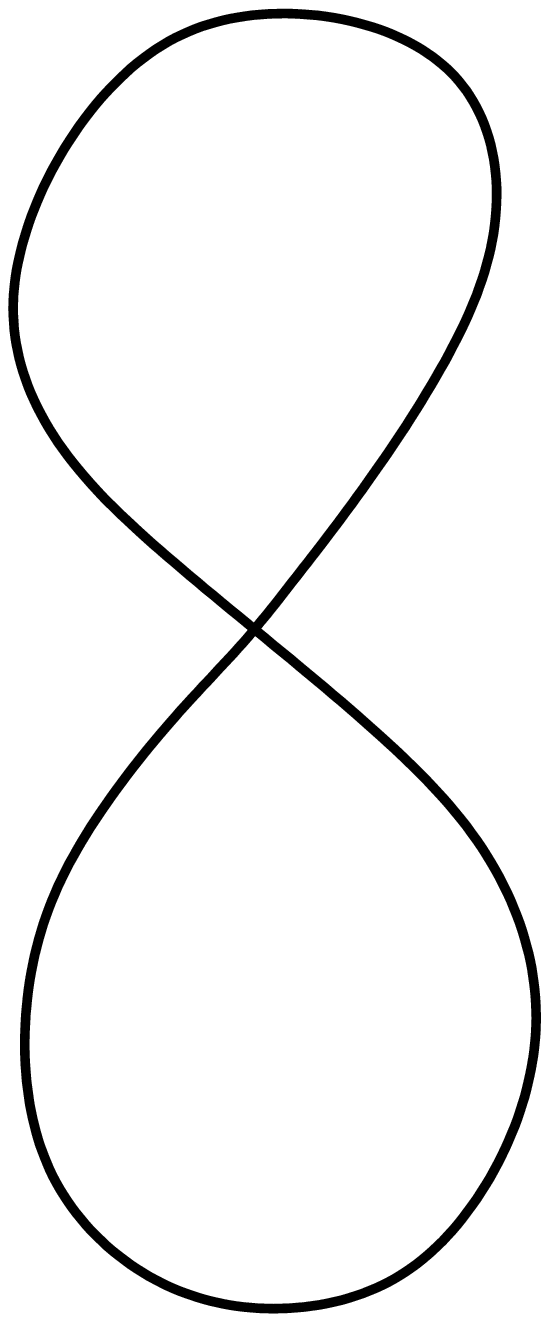}
    \end{array}} 
        = \begin{array}{c}
  	\mbox{\includegraphics[width=1cm]{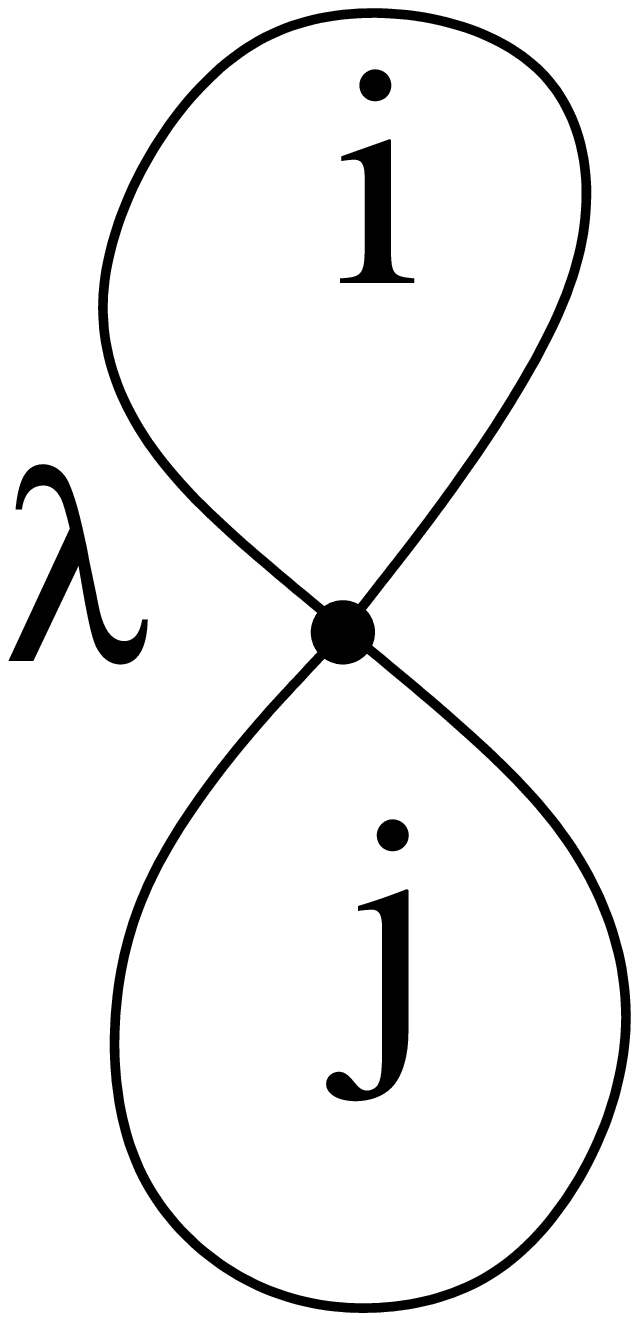}}
    \end{array},
\end{eqnarray}

\begin{eqnarray}
\nonumber &&\left|ijkl, \lambda \gamma \right>_{\begin{array}{c}
	\includegraphics[width=.65cm]{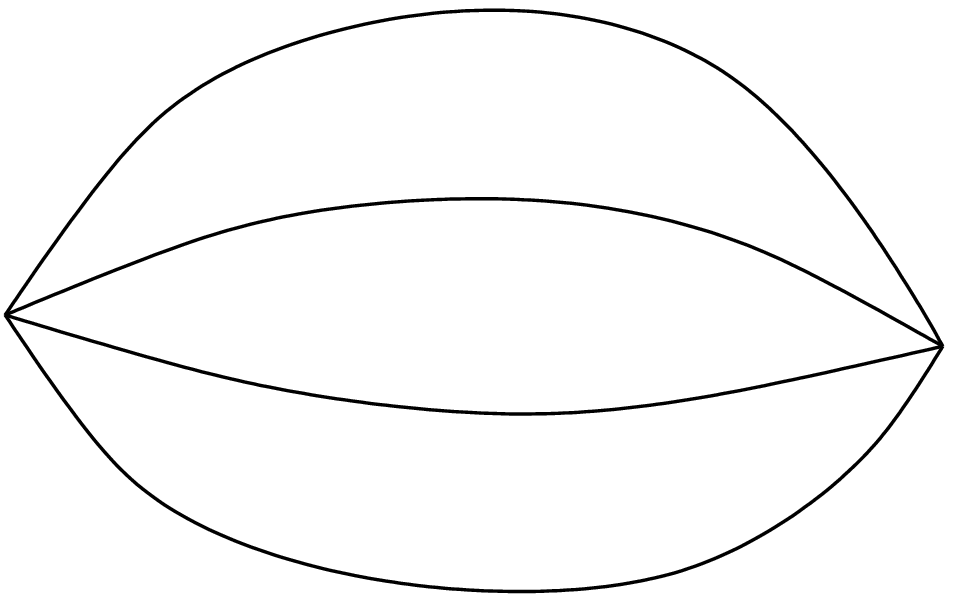}
    \end{array} } 
        =\begin{array}{c}
  	\mbox{\includegraphics[width=2.6cm]{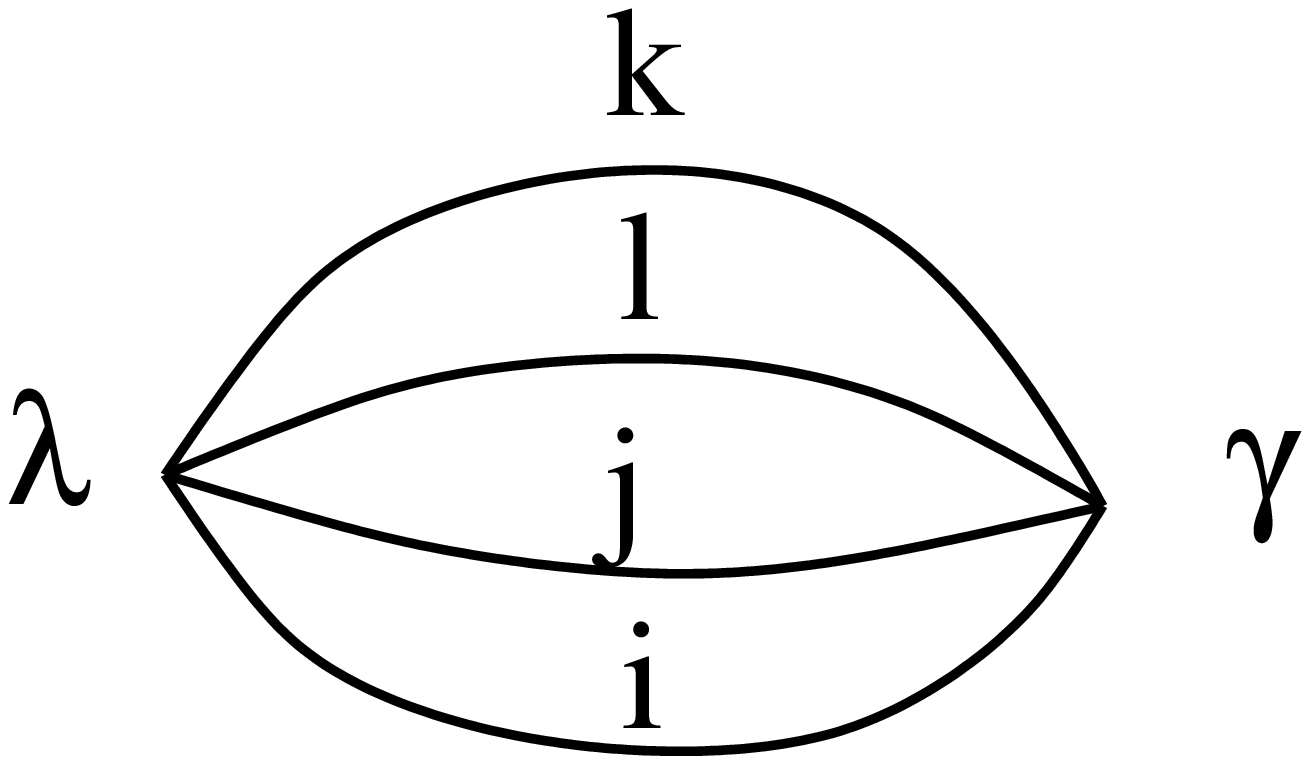}}
    \end{array} \ \ {\rm for}\ \  \lambda \not= \gamma,
\end{eqnarray}
\begin{eqnarray}
\nonumber &&\left|ijkl, \lambda \gamma\right>_{\begin{array}{c}
	\includegraphics[width=.65cm]{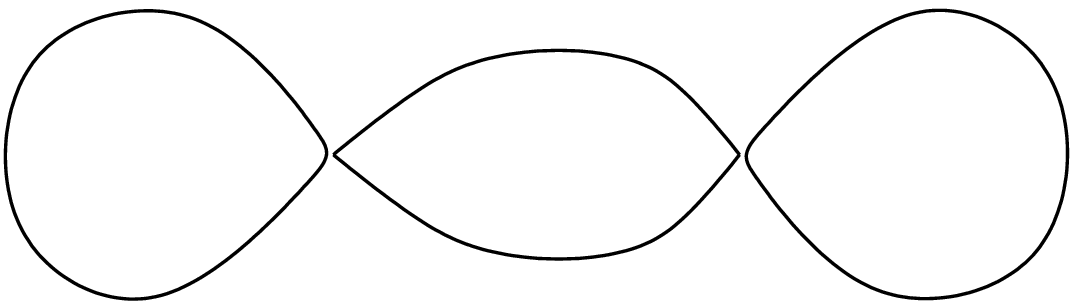}
    \end{array}} 
        =
    \begin{array}{c}
  	\mbox{\includegraphics[width=3cm]{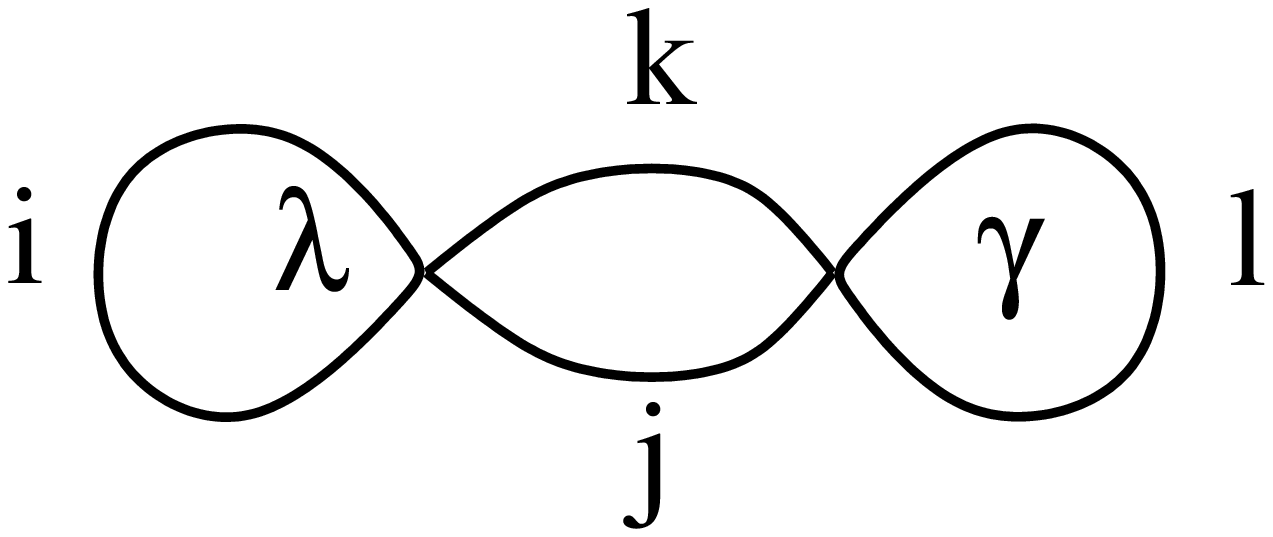}}
    \end{array} \ \ {\rm for}\ \  \lambda \not= \gamma \ \ {\rm or} \ \ i\not=l,
\end{eqnarray}
\begin{eqnarray}
\nonumber &&\left| ijklmn,\lambda \gamma \delta \right>_{\begin{array}{c}   
	\includegraphics[width=.25cm]{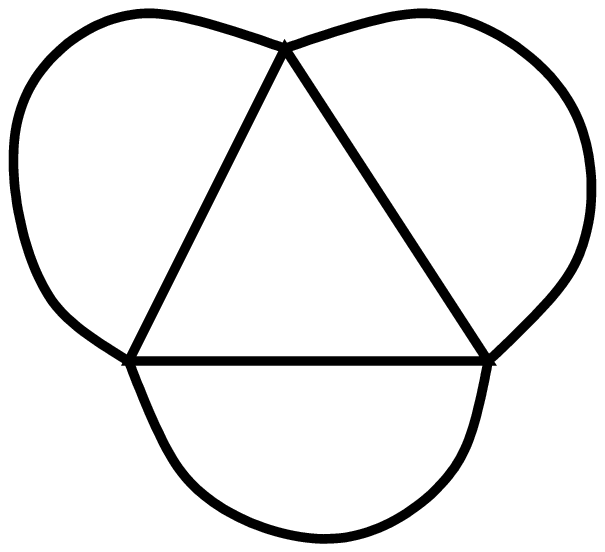}                
\end{array}} =
    \begin{array}{c}
  	\mbox{\includegraphics[width=2.5cm]{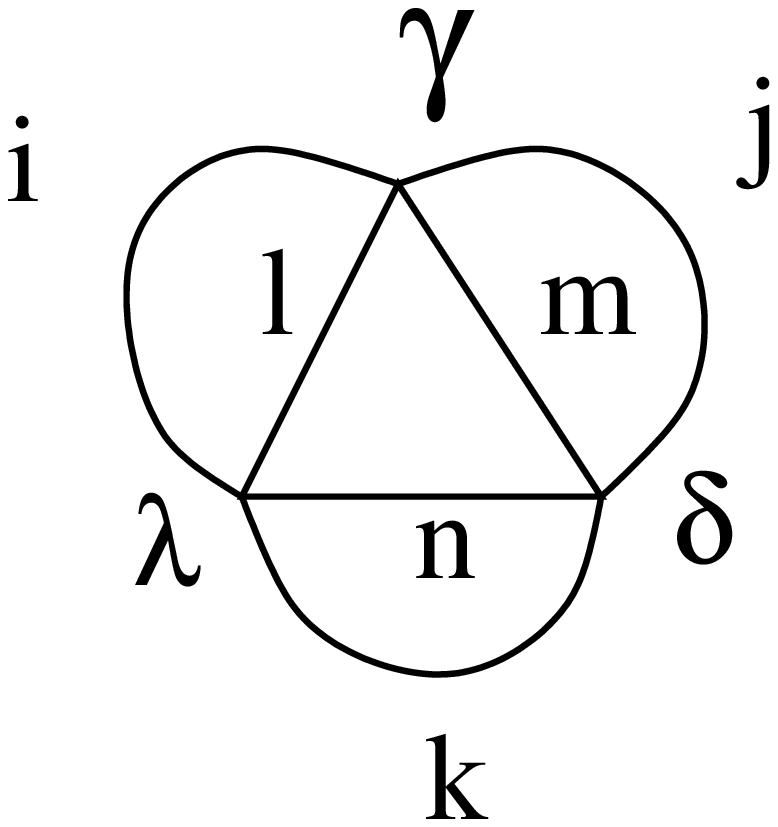}}
    \end{array} .
\end{eqnarray}
The states which interfere with the vacuum are those for which there are
closed bubble diagrams from the given spin network to `nothing'. In those
orthonormal states in the physical state can constructed by simply subtracting
the vacuum part using the standard Gram-Schmidt procedure. For example
\begin{eqnarray}
\nonumber &&\left|ijkl, \lambda \right>_{\begin{array}{c}
	\includegraphics[width=.65cm]{2.eps}
    \end{array} } 
        =-1 +
    \begin{array}{c}
  	\mbox{\includegraphics[width=2.8cm]{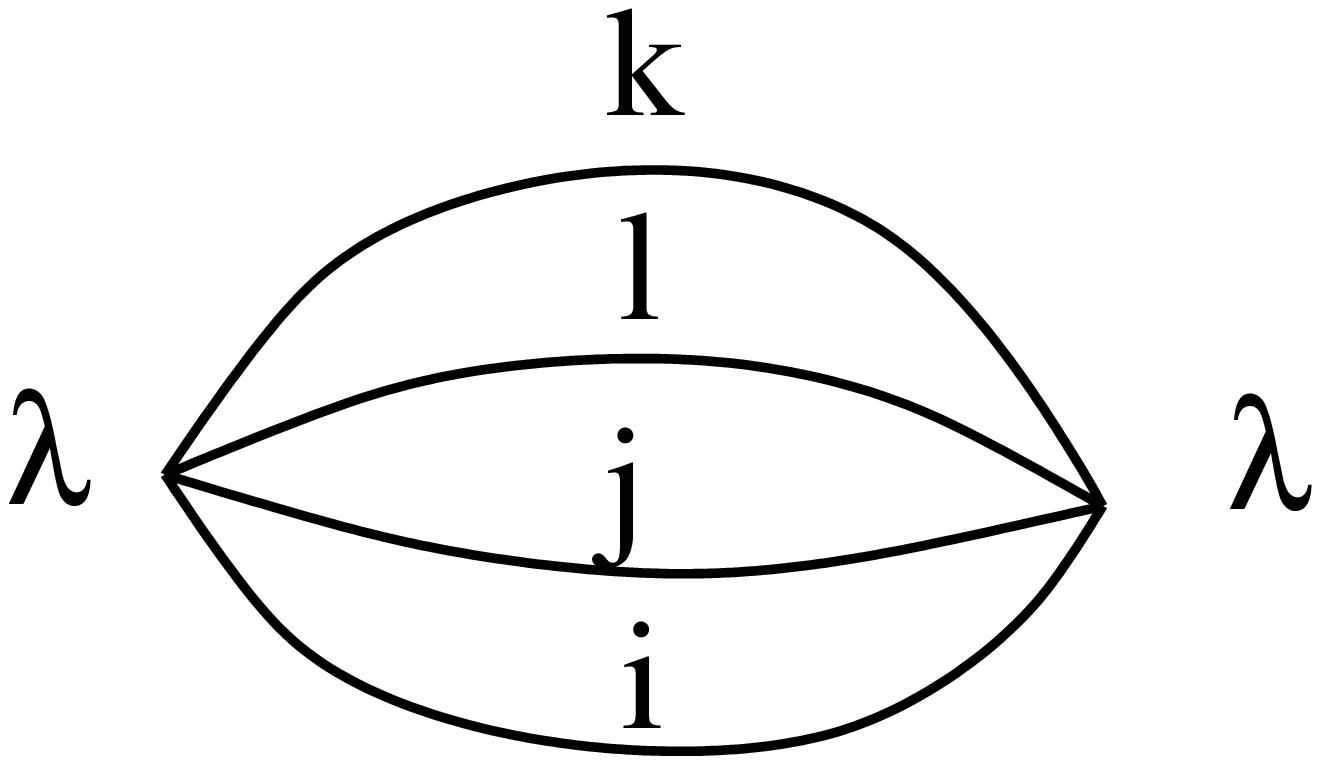}}
    \end{array},
\end{eqnarray}
\begin{eqnarray}
\nonumber &&\left|ijk, \lambda \right>_{\begin{array}{c}
	\includegraphics[width=.65cm]{2bis.eps}
    \end{array} } 
        =-1 +
    \begin{array}{c}
  	\mbox{\includegraphics[width=3cm]{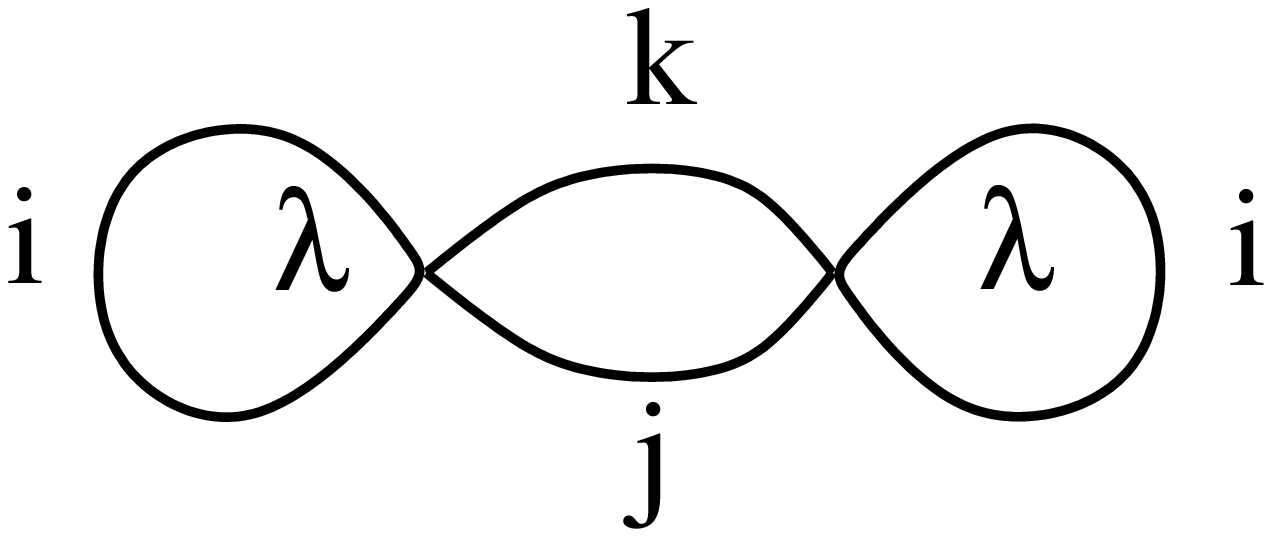}}
    \end{array},
\end{eqnarray}
\begin{eqnarray}
\nonumber &&\left|ij,\lambda \right>_{\begin{array}{c}
	\includegraphics[width=.25cm]{1.eps}
    \end{array}
\begin{array}{c}
\includegraphics[width=.25cm]{1.eps}
    \end{array}}
        =-1 +
    \begin{array}{c}
  	\mbox{\includegraphics[width=1cm]{1b.eps}}
    \end{array}
\begin{array}{c}
\includegraphics[width=1cm]{1b.eps}
    \end{array},
\end{eqnarray}
and so on. Other states can be turn out to be just the tensor product
of the previous ones, namely
\begin{eqnarray}
\nonumber &&\left|ij kl mn, \lambda \delta  \right>_{\begin{array}{c}
	\includegraphics[width=.25cm]{1.eps}
    \end{array} \begin{array}{c}
	\includegraphics[width=.25cm]{2.eps}
    \end{array}}
        = - \begin{array}{c}
	\includegraphics[width=1cm]{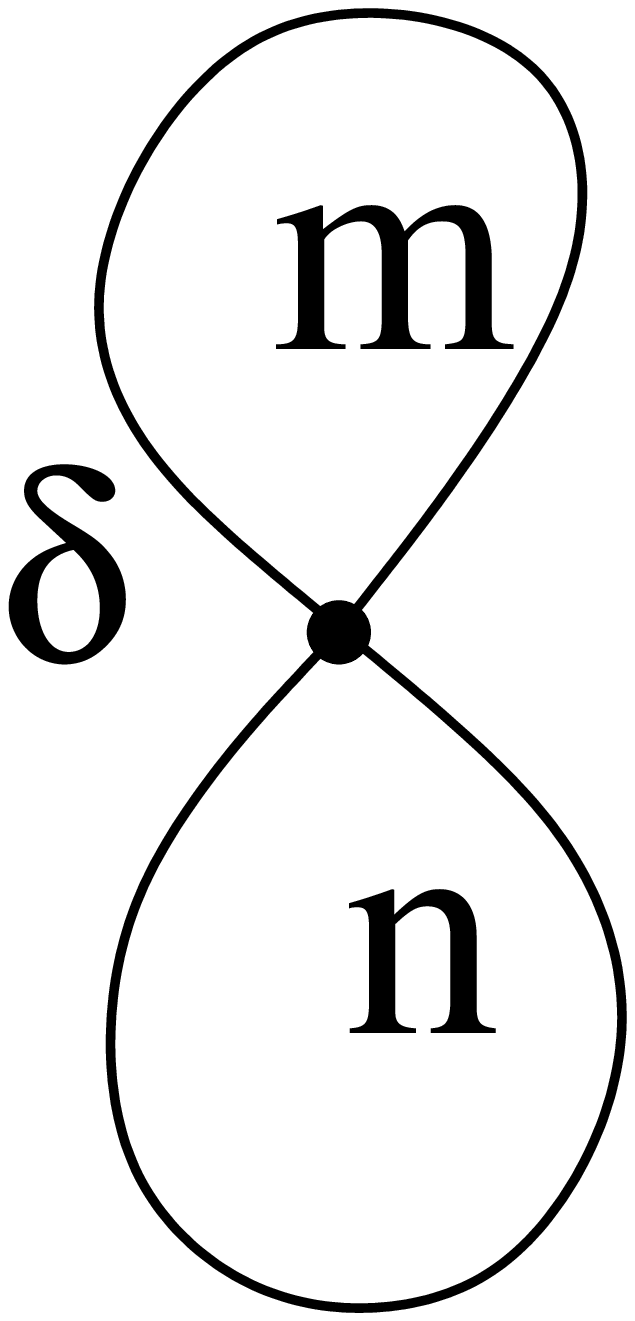}
    \end{array} +
    \begin{array}{c}
  	\mbox{\includegraphics[width=1cm]{1bb.eps}}
    \end{array}\begin{array}{c}
	\includegraphics[width=3cm]{2ba.eps}
    \end{array}, 
\end{eqnarray}
\begin{eqnarray}
\nonumber &&\left| ij k mn, \lambda  \delta  \right>_{\begin{array}{c}
	\includegraphics[width=.25cm]{1.eps}
    \end{array} \begin{array}{c}
	\includegraphics[width=.25cm]{2bis.eps}
    \end{array}}
        =-\begin{array}{c}
	\includegraphics[width=1cm]{1bb.eps}
    \end{array} +
    \begin{array}{c}
  	\mbox{\includegraphics[width=1cm]{1bb.eps}}
    \end{array}\begin{array}{c}
	\includegraphics[width=3cm]{dos.eps}
    \end{array}, 
\end{eqnarray}
\begin{eqnarray}
\nonumber &&\left| ij  mn,\lambda  \delta   \right>_{\begin{array}{c}
	\includegraphics[width=.25cm]{1.eps}
    \end{array} \begin{array}{c}
	\includegraphics[width=.25cm]{1.eps}
    \end{array}\begin{array}{c}
	\includegraphics[width=.25cm]{1.eps}
    \end{array}} 
        =- \begin{array}{c}
	\includegraphics[width=1cm]{1bb.eps}
    \end{array} + \begin{array}{c}
  	\mbox{\includegraphics[width=1cm]{1b.eps}}
    \end{array}\begin{array}{c}
	\includegraphics[width=1cm]{1b.eps}
    \end{array}\begin{array}{c}
	\includegraphics[width=1cm]{1bb.eps}
    \end{array},
\end{eqnarray}
and so on.  The procedure can be clearly continued
to construct an orthonormal basis of ${\cal H}_{ph}.$ 

\section{Complex $P$}\label{openissues} 

An important ingredient in the construction above is the 
assumption that $P$ is real, equation 
(\ref{eq:reality}).  This assumption greatly simplifies 
the construction, allowing a simple definition of the 
$\phi_{s}$ operators. Here we discuss the meaning of 
this assumption and the extension of the formalism to 
the case in which $P$ is non-real, or ``complex''.
\begin{figure}[h]
	\centering \vskip2cm 
\includegraphics[width=2.5cm]{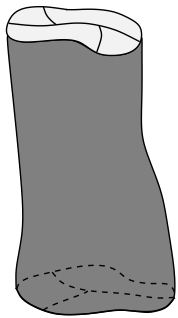}
	\caption{The three-geometry to three-geometry transition 
	 amplitude $W(s,s')$ between two connected 
	 three-geometries.}
	 \label{uno} 
\end{figure}
\begin{figure}[h]
	\centering \vskip2cm 
\includegraphics[width=4cm]{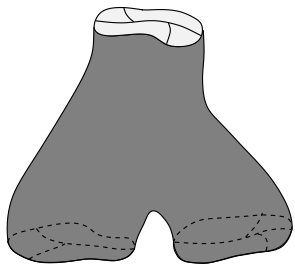}
	\caption{The transition 
	 amplitude between a doubly connected and a connected 
	 three-geometry.}
	 \label{due} 
\end{figure}
\begin{figure}[h]\vskip2cm 
\centering

$\begin{array}{c} \mbox{\includegraphics[width=4cm]{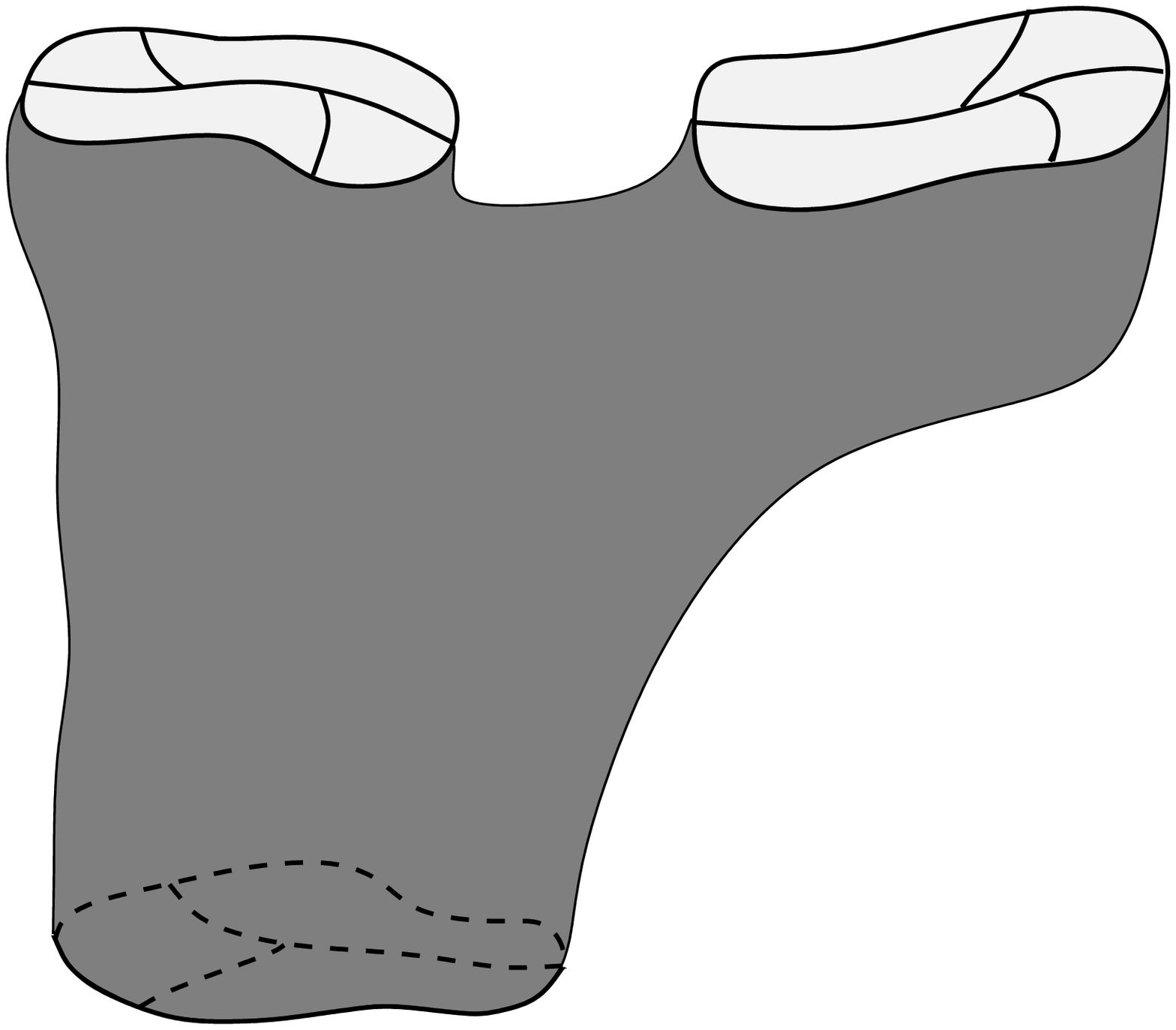}}\end{array}$\ \ 
= 
\ \ $\begin{array}{c} \includegraphics[width=4cm]{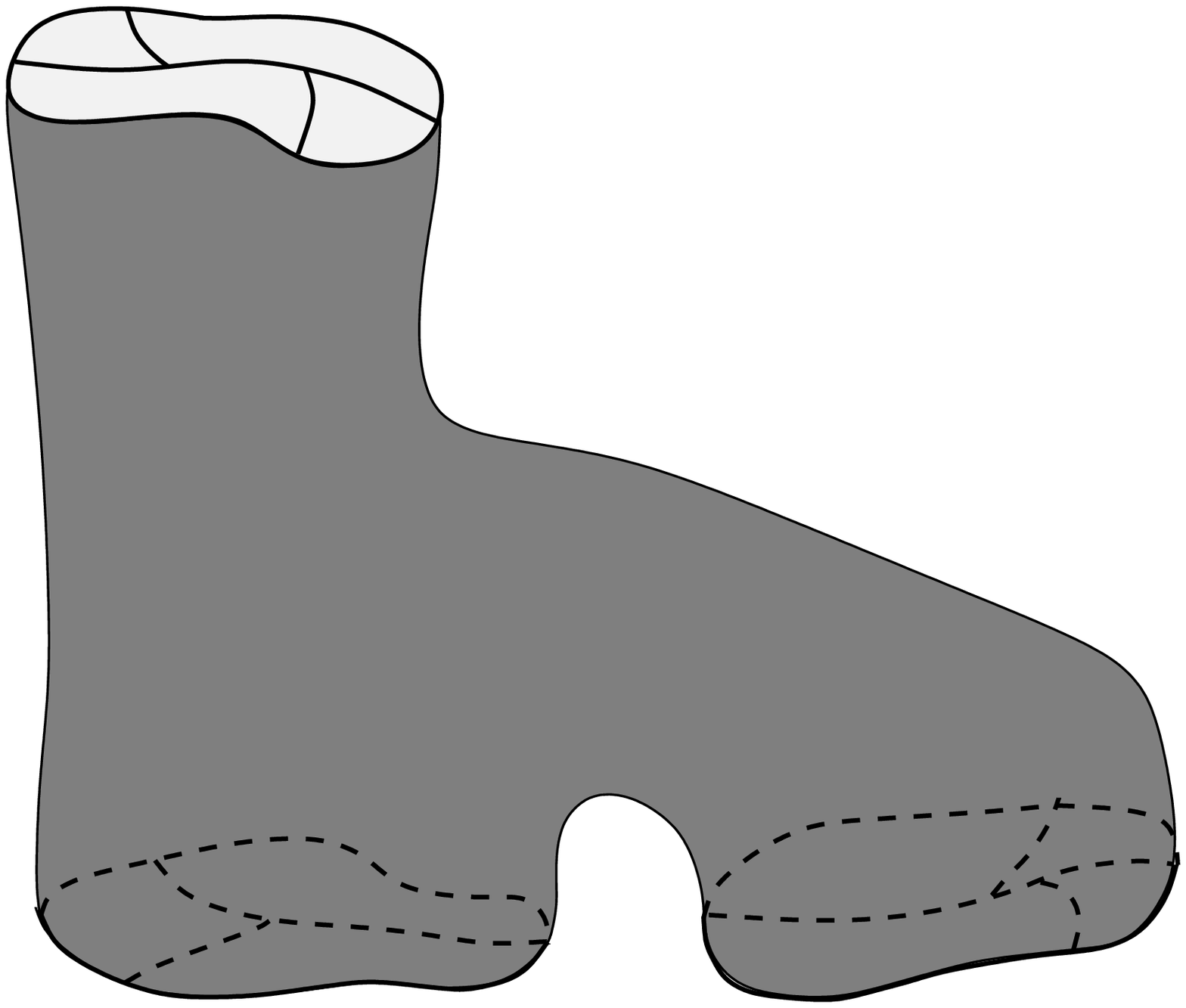} \end{array}$
	\caption{The reality condition on $P$.}
	\label{tre}
\end{figure}

To clarify the meaning of the realty condition, let us represent
graphically the 2-net function $W(s,s')$ as in Figure 1, when $s$ and
$s'$ are connected.  If $s'$ is formed by two connected components
$s_{1}$ and $s_{2}$, we represent it as in Figure 2.  Then the reality
of $P$ is expressed by the equation in Figure 3.  That is, it
represents an a priori indistinguibility between past and future
boundaries of spacetime.  This property is strictly connected with
crossing symmetry \cite{CarloMike}, which is essentially the analogous
property at the level of the Hamiltonian constraint.  The property is
natural from the perspective of Atiyah's topological quantum field
theory axiomatic framework \cite{atihay}.  It is perhaps natural to
expect this property for Euclidean quantum gravity.  Wether we should
expect Lorentzian quantum gravity to have the same property, on the
other hand, is not clear to us.  On the one hand, the causal structure
of the Lorentzian four-geometries seems to suggest that one should
distinguish past and future boundaries.  Notice also that
(\ref{eq:reality}) implies that the transition amplitudes $W(s,s')$
are real, because
\begin{equation}
	W(s,s')=
        \langle s P s'\rangle=
        \langle 0 P s'\cup s \rangle=
        \langle s' P s \rangle=
        \overline{\langle s P s' \rangle} =
        \overline{W(s,s')} ,
	\label{real W}
\end{equation} 
which would prevent quantum mechanical interference 
between the $| s \rangle$ basis states (but not between 
generic states).  On the other hand, however, temporal 
relations between boundaries may be induced a posteriori 
by the dynamics, instead of being a priori given in the 
structure of the formalism itself.

If we drop the reality condition on $P$, the main 
difficulty is that the definition (\ref{phifull}) of the 
field operator becomes inconsistent.  However, can still 
retain a (partial) characterization of the field operator 
$\phi_{s}$ by requiring only 
\begin{equation} 
\hat\phi_{s}|0 \rangle_{ph}=|s \rangle_{ph}.  
\label{phi0}
\end{equation}
This is certainly consistent. Notice that in 
general we have then 
\begin{equation}
    \phi^{\dagger}_{s}\ne\phi_{s}. 
    \label{eq:notequal}
\end{equation}
And the 2-net function is now given by
\begin{equation}
    W(s,s') = {}_{ph}\langle 0|\hat\phi^{\dagger}_{s} \hat\phi_{s'}|0
    \rangle_{ph}.
\label{sistar}
\end{equation}
That is, we have to add the adjoint operation to 
equation (\ref{nostar}).

The relevant abstract $C^{*}$-algebra has now a non 
trivial star operation, different from $s^{*}=s$.  If 
$\hat\phi^{\dagger}_{s}$ and $\hat\phi_{s'}$ are 
independent, the $C^{*}$-algebra is generated by 
products of $s$'s and $s^{*}$'s, and 
 \begin{equation}
         W(s,s') = W(s^{*}\cup s').
 	\label{Wstar}
 \end{equation} 

Starting from the field theory, we may generate a $W$ 
functional on the complex algebra $\cal A$ by using a 
complex field, instead than a real one.  The resulting 
structure will be explored in detail elsewhere.

A strictly related problem is whether the $n$-net $W$ functions should
be thought as analogous to the Wightman distributions (the vacuum
expectation values of products of field operators), to the Feynman
distributions (the vacuum expectation values of time ordered products
of field operators) or rather to the Schwinger functions (the
appropriate analytic continuations of the Wightman distributions to
imaginary time).  We recall that on Minkowski space one can directly
apply the GNS reconstruction theorem to the Wightman distributions. 
On the other hand, one obtains directly the Feynman distributions as
functional integrals of products of fields (with suitable
``prescriptions'' at the poles), while one can obtain the Schwinger
functions as momenta of a well defined stochastic process \cite{W}. 
The Osterwalder-Schrader reconstruction theorem \cite{OS} that allows
the reconstruction of the Wightman distributions from the Schwinger
functions requires a duality ``star'' operation to be defined,
corresponding to the inversion of the time variable.  Presumably, the
distinction between these different families of $n$-point functions
makes no sense in the generally covariant context.  The peculiar
analytic structure of the $n$-point functions of field theory on
Minkowski space is a consequence of the positivity of the energy (the
Fourier transform of a function with support on positive numbers is
analytic in the upper complex plane.), while there is no notion of
positivity of the energy in quantum gravity -- indeed, there is no
notion of energy at all--, and one should be careful in trying to
generalize standard quantum field theoretical prejudices to the
generally covariant context.

\section{Conclusion}

We have studied the family of quantities $W(s)$, which we propose as
main physical observables of a quantum theory of gravity.  We have
proposed a general framework, based on these quantities, that ties the
canonical (loop) and the covariant (spin foam) approaches to quantum
gravity.  The connection between the two formalisms is provided by the
GNS reconstruction theorem, and parallels the connection between the
Hilbert space and the functional formulations of conventional quantum
field theory, which one obtains from the properties of the $n$-point
functions.

Many issues deserve to be clarified.  Among these are the reality of
$P$ and the complex $\cal A$ algebra; the locality property of $W(s)$
mentioned at the end of Section \ref{Wfunctions}; and the connection
between $W(s)$ and the $S$ matrix when spacetime admits asymptotic
regions.  An explicit construction of the $C^*$ algebra and the its
GNS construction in the case of the 2 dimensional theory \cite{2dft}
would also be of great interest and presumably not too hard to do.

As mentioned in the introduction, an independent but related approach
has recently appeared in Ref.\ \cite{mikovic} by Alexandar Mikovic. 
Mikovic studies the field theories of a group, and constructs a the
Hilbert space of the theory --in fact, a Fock space-- directly from
the theory's field operators, instead of using the intermediary step
of the transition functions as we do here.  He observes, as we do
here, that this Hilbert space has a natural basis of spin network
states, where the spin network correspond to a triangulation of the
spacetime boundary, and that in perturbation theory the transition
amplitudes between spin network amplitudes are given by the state-sum
amplitudes for triangulated manifolds with boundaries.  This structure
parallels what is done here.  It may be useful on the other hand, to
emphasize a difference in attitude: Mikovic sees the discrete aspects
of the geometry as a starting postulate; here, on the contrary, we
view them as a result of the canonical quantization of the continuum
theory. 

In addition, Mikovic considers evolution in a fixed number of
simplices, and proposes to interpret the number of simplices between
two spin networks as a time parameter.  Transition amplitudes in this
time are then finite.  The same idea of fixing the number of Planck
steps to get a time variable and finite transition amplitudes has been
explored by Markopoulou and Smolin \cite{lee}, and can be traced to
Sorkin's suggestion of using the 4d volume as a natural time variable
\cite{sorkin}.  See also Teitelboim's \cite{teitelboim}.  The resulting
transition amplitudes are analogous to Feynman's proper time
propagator $P(x,y;\tau)$ obtained integrating over all paths in
Minkowski space that go from $x$ to $y$ in the proper time $\tau$. 
The physical propagator $P(x,y)$ is obtained from $P(x,y;\tau)$ by
integrating over $\tau$.  Indeed, it is this integration that amounts
to impose the Hamiltonian constraint.  As far as we understand, the
unintegrated proper time propagator does not have a simple physical
interpretation in the context of the dynamics of the particle.  The
analogy suggests that in the gravitational context we cannot confine
ourselves to the transition amplitudes in a fixed number of Planck
steps, but the issue deserves to better understood.

\end{document}